\documentclass[conference]{IEEEtran}
\IEEEoverridecommandlockouts

\usepackage{cite}
\usepackage{hyperref}
\usepackage{amsmath,amssymb,amsfonts}
\usepackage{algorithmic}
\usepackage{graphicx}
\usepackage{textcomp}
\usepackage{xcolor}
\usepackage{multirow}
\usepackage{multicol}
\usepackage{booktabs}
\usepackage{microtype}
\AtBeginDocument{\setlength{\abovedisplayskip}{4pt}\setlength{\belowdisplayskip}{4pt}\setlength{\abovedisplayshortskip}{2pt}\setlength{\belowdisplayshortskip}{2pt}}
\def\BibTeX{{\rm B\kern-.05em{\sc i\kern-.025em b}\kern-.08em
    T\kern-.1667em\lower.7ex\hbox{E}\kern-.125emX}}
\begin{document}

\title{LACE: Layer-Wise Compression for Dynamic Frame Rate Codecs
}


\author{\IEEEauthorblockN{Thanapat Trachu, Samuele Cornell,  William Chen, Shinji Watanabe}
\IEEEauthorblockA{\textit{Language Technologies Institute, Carnegie Mellon University}, Pittsburgh, USA \\
\{ttrachu, scornell, wc4, swatanab\}@andrew.cmu.edu}
}

\maketitle

\begin{abstract}
Neural audio codecs are a key component in speech language modeling. However, their high frame rates lead to long sequence lengths, increasing computational costs. Dynamic frame rate codecs mitigate this by reducing the effective frame rate using a compression step to merge multiple frames together. However, most prior methods either operate on single-codebook codecs or apply a single compression step before multi-layer quantization. This forces all quantization layers to share the same segmentation boundaries, despite the residual embeddings at different quantization layers exhibiting different rates of change over time.
We propose LACE (Layer-Adaptive Codec Encoding), a dynamic frame rate codec that applies an independent compression step at each quantization layer, enabling layer-specific segmentation boundaries. To use LACE tokens in downstream text-to-speech (TTS), we further introduce union alignment and boundary anchor mechanisms to make durations consistent across layers while preserving compression benefits. Experiments on LibriTTS show that LACE offers a better rate-quality tradeoff than prior dynamic frame rate methods on the reconstruction task and improves TTS inference efficiency while maintaining competitive synthesis quality. Our code is released as part of the ESPnet3 codec recipe.
\end{abstract}

\begin{IEEEkeywords}
Audio Codec, Text-to-Speech, Dynamic Frame Rate Codec, Compression
\end{IEEEkeywords}

\section{Introduction}
Neural audio codecs have emerged as an important component for speech language modeling. When used as speech tokenizers, they convert continuous speech into discrete codes, allowing speech generation~\cite{wang2025neuralcodeclanguagemodels, chen2025valle, voicecraft} and speech language models~\cite{fastlongspeech, tian2025espnetspeechlm, yang2024uniaudio} to be formulated as next-token prediction problems similar to those used in large language models. However, general neural audio codecs \cite{dac_codec, soundstream_codec, encodec_codec} typically operate at high frame rates, producing long sequences of discrete codes.

Long sequences of discrete codes introduce two limitations. First, long sequences increase the computational cost, as the complexity of transformer-based models tends to scale with sequence length \cite{audiolm}. Second, the length mismatch between speech and text tokens can degrade the performance of speech language models, since substantially longer speech sequences make syntactic and semantic modeling more difficult, as shown in \cite{wang2024whyspeech}.

These limitations motivate dynamic frame rate codecs \cite{zhang25k_interspeech, zheng2025sayless, li2025flexicodec, codecslime}, which represent speech as discrete codes paired with durations. A compression model produces \emph{segmentation boundaries}, which are frame positions where the frame-level embeddings are partitioned. These boundaries divide the embeddings into variable-length segments, and each segment is represented by one code and its duration. Since each code now corresponds to a segment rather than a single frame, the frame rate is measured as the number of segments per second instead of the number of frames per second. We refer to this as the \emph{effective frame rate}. For example, if one second of audio with 75 frame-level embeddings is compressed into 25 segments, the effective frame rate is 25 Hz, instead of 75 Hz. By assigning longer durations to regions with lower information density, dynamic frame rate codecs use fewer codes to represent speech than general codecs.

Most existing dynamic frame rate codecs either operate on single-codebook codec models \cite{codecslime, zheng2025sayless}, or apply a single compression step before multi-layer quantization \cite{li2025flexicodec, carve, qian2026arbitrarily}. In both cases, all quantization layers are constrained to share the same segmentation boundaries. This constraint may be suboptimal for multi-codebook codecs due to the hierarchical nature of their quantization process. In residual vector quantization (RVQ)~\cite{rvq}, the first layer quantizes the input directly, while each subsequent layer quantizes the residual embeddings from the previous layer. Consequently, earlier layers capture the main signal structure, whereas deeper layers encode finer details. As a result, representations at different layers may change at different rates over time, suggesting that each layer requires its own segmentation boundaries. This intuition is supported by our analysis in Section~\ref{sec:analysis_cosine}, which shows that deeper-layer residual embeddings change more rapidly than earlier-layer residual embeddings. This is further supported by the findings of SNAC~\cite{siuzdak2024snac}, which uses different frame rates across quantization layers, although it is not a dynamic frame rate codec because its frame rates are fixed.

Motivated by these findings, we propose LACE (\textbf{L}ayer-\textbf{A}daptive \textbf{C}odec \textbf{E}ncoding), a dynamic frame rate codec that independently applies a compression step at each quantization layer. This allows each layer to choose its own segmentation boundaries.
To adapt LACE to the common speech language model framework, we propose union alignment, which constructs shared segmentation boundaries by taking the union across layers. Since union alignment can increase the effective frame rate, we further propose boundary anchor, which constrains deeper layers to reuse boundaries introduced by earlier layers. Our contributions are summarized as follows:
\begin{itemize}\setlength{\itemsep}{0pt}\setlength{\parskip}{0pt}
    \item We propose LACE, a dynamic frame rate codec that applies an independent compression step at each quantization layer.
    \item We propose union alignment and boundary anchor to resolve duration inconsistency across layers, enabling LACE tokens to be used in downstream TTS training.
    \item We provide a theoretical analysis showing that LACE achieves a lower upper bound on the expected quantization error than single compression.
    \item Experiments on LibriTTS~\cite{zen19_interspeech} show that LACE offers a better rate-quality tradeoff than prior dynamic frame rate methods on the reconstruction task and improves TTS inference efficiency while maintaining competitive synthesis quality.
\end{itemize}

\section{Background}
\subsection{Neural Audio Codec Preliminaries}
\label{sec:preliminaries}
A neural audio codec consists of an encoder, a quantization module, and a decoder. The encoder maps the raw waveform into frame-level embeddings $\mathbf{H} \in \mathbb{R}^{T \times D}$, where $T$ is the number of frames and $D$ is the embedding dimension. The quantization module discretizes $\mathbf{H}$ into frame-level codes $\mathbf{Z} \in [1,V]^{T \times C}$, where each frame is represented by $C$ codes from a codebook of size $V$. The decoder then reconstructs the waveform from these codes.

A commonly used quantization method is RVQ. RVQ discretizes $\mathbf{H}$ using $C$ quantization layers, where each layer $l$ contains a codebook $\mathbf{Q}^l \in \mathbb{R}^{V \times D}$. Here, $\mathbf{q}_k^l \in \mathbb{R}^D$ denotes the $k$-th codeword of $\mathbf{Q}^l$. RVQ performs quantization sequentially across quantization layers. At layer $l$, the nearest codeword is selected from the corresponding codebook $\mathbf{Q}^l$ as $z_i^l = \arg\min_{k \in [1,V]} \| \mathbf{r}_i^l - \mathbf{q}_k^l \|_2^2$, where $z_i^l \in [1,V]$ and $\mathbf{r}_i^l \in \mathbb{R}^D$ denote a code and a residual embedding at frame $i$ and quantization layer $l$, respectively. The residual embedding is updated recursively as $\mathbf{r}_i^{l+1} = \mathbf{r}_i^l - \mathbf{q}_{z_i^l}^l$, with $\mathbf{r}_i^1 = \mathbf{h}_i$, where $\mathbf{h}_i \in \mathbb{R}^D$ is the $i$-th frame-level embedding in $\mathbf{H}$. We define the stacked frame-level residual embeddings of layer $l$ as $\mathbf{R}^{l} \in \mathbb{R}^{T\times D}$.

\subsection{Compression in Dynamic Frame-Rate Codecs}
\label{sec:compression_background}

The encoder of a neural audio codec produces frame-level embeddings $\mathbf{H}$ at a fixed codec frame rate, yielding $T$ frames. A compression model groups these $T$ frames into $M$ variable-length segments. Each segment is represented by $C$ codes and an integer duration.

Formally, the compression model determines segmentation boundaries $\mathcal{B} = \{b_0, b_1, \dots, b_M\}$, a sorted set with $b_0=0$, $b_M=T$, and $b_m \in [1,T-1]$. The $m$-th segment spans the frame interval $(b_{m-1}, b_m]$. Some methods control $M$ explicitly using a target compression rate $\gamma \in (0,1]$, yielding $M = \lfloor \gamma T \rfloor$, while others determine $M$ implicitly through a threshold $\tau$. The frame-level embeddings within each segment are averaged to produce segment-level embeddings $\bar{\mathbf{H}} \in \mathbb{R}^{M \times D}$. The segment-level embeddings are paired with the durations $\mathbf{d} \in \mathbb{N}^M$, where the duration of the $m$-th segment $d_m$ is $b_m-b_{m-1}$. The segment-level embeddings $\bar{\mathbf{H}}$ are quantized using RVQ, as described in Section~\ref{sec:preliminaries}, producing segment-level codes $\bar{\mathbf{Z}} \in [1,V]^{M \times C}$. This (code, duration) format resembles the classical run-length encoding scheme~\cite{golomb1966runlength}. During reconstruction, $\bar{\mathbf{Z}}$ is repeated according to $\mathbf{d}$ to produce frame-level codes $\hat{\mathbf{Z}} \in [1,V]^{T \times C}$. This restores the frame rate expected by the decoder. The three compression models below share this pipeline and differ only in how they derive $\mathcal{B}$.

\paragraph{Dynamic Programming (DP)-Based Compression.}
CodecSlime~\cite{codecslime} determines $\mathcal{B}$ by minimizing the total reconstruction error between the original frame-level embeddings $\mathbf{H}$ and the segment-level embeddings $\bar{\mathbf{H}}$. For a segment of $s$ consecutive frames ending at frame $j$, the segment-level embedding is defined as the mean of the frames within the segment: $\bar{\mathbf{h}}_{j,s} = \frac{1}{s} \sum_{t=j-s+1}^{j} \mathbf{h}_t$.
The reconstruction loss for this segment is $\ell(j,s) = \sum_{t=j-s+1}^{j} \| \mathbf{h}_t - \bar{\mathbf{h}}_{j,s} \|_2^2$. Each segment is constrained to contain at most $U$ frames. Let $f[j,n]$ denote the minimum total reconstruction loss when the first $j$ frames are compressed into $n$ segments. The DP objective is defined recursively as:
\begin{align} \label{eq:dp_obj}
    f[j,n] = \min_{1 \leq s \leq U} \left\{ f[j-s,\, n-1] + \ell(j,s)
    \right\},
\end{align}
with $f[0,0] = 0$. At each state $(j,n)$, we record the segment length $s^*[j,n]$ that achieves the minimum in Equation \eqref{eq:dp_obj}. Given the number of segments $M$, we start from the last boundary $b_M=T$ and recover each preceding boundary as $b_{n-1} = b_n - s^*[b_n,n]$ for $n = M, M{-}1, \dots, 1$, which terminates at $b_0=0$ and yields the segmentation boundaries $\mathcal{B}$.

\paragraph{Cosine Similarity-Based Compression.}
FlexiCodec~\cite{li2025flexicodec} determines $\mathcal{B}$ by thresholding the similarity between adjacent frame-level embeddings. Let $\sigma_t = \cos(\mathbf{h}_t, \mathbf{h}_{t+1})$ denote the similarity between frames $t$ and $t+1$. A boundary is placed wherever the similarity drops below a threshold $\tau$, yielding $\mathcal{B} = \{0,T\} \cup \{\, t : \sigma_t < \tau \,\}$.
The number of segments $M$ is therefore not fixed in advance but controlled implicitly by the threshold $\tau$, unlike the DP-based compression model.

\paragraph{Density Peak Clustering-Based Compression.}
VARSTok~\cite{zheng2025sayless} derives $\mathcal{B}$ using density peak clustering. For each frame $t$, a local density $\rho_t$ is computed from its $\kappa$ nearest neighbors, and a peak distance $\delta_t$ is the temporal distance to the nearest frame with higher density:
\begin{align}
    \rho_t = \frac{1}{\kappa} \sum_{j \in \text{KNN}(t)}
    \phi(\mathbf{h}_t, \mathbf{h}_j), \qquad
    \delta_t = \min_{j:\rho_j > \rho_t} |j-t|,
\end{align}
where $\phi(\mathbf{h}_i,\mathbf{h}_j) = (1+\langle \mathbf{h}_i,\mathbf{h}_j\rangle)/2$ and $\text{KNN}(t)$ denotes the set of $\kappa$ nearest neighbors of frame $t$. Frames with a high peak score $p_t = \rho_t \delta_t$ are selected as segment centers. Each center is greedily expanded to adjacent unassigned frames that satisfy $\phi(\mathbf{h}_{i^*}, \mathbf{h}_t) - \beta p_t > \tau$, up to a maximum of $U$ frames. This process repeats until all frames are assigned to segments. The segment endpoints are then sorted to form $\mathcal{B}$. The number of segments $M$ is controlled implicitly by $\tau$. We use $\kappa=5$, $\beta=0.2$, and $U=4$, following the original paper.

All of these prior works force all quantization layers to share the same segmentation boundaries. In contrast, LACE applies an independent compression step at each quantization layer, enabling layer-specific segmentation boundaries. Because LACE is agnostic to the choice of compression model, any of these methods can be used as the compression step within the proposed framework.

\section{Method}
We propose LACE, a dynamic frame rate codec that applies an independent compression step at each quantization layer, allowing each layer to choose its own segmentation boundaries. We then introduce two mechanisms, union alignment and boundary anchor, to make the duration consistent for downstream TTS training.

\subsection{Layer-Wise Compression} \label{sec:multi_compression}

\begin{figure}[t]
    \centering
    \includegraphics[width=.28\textwidth]{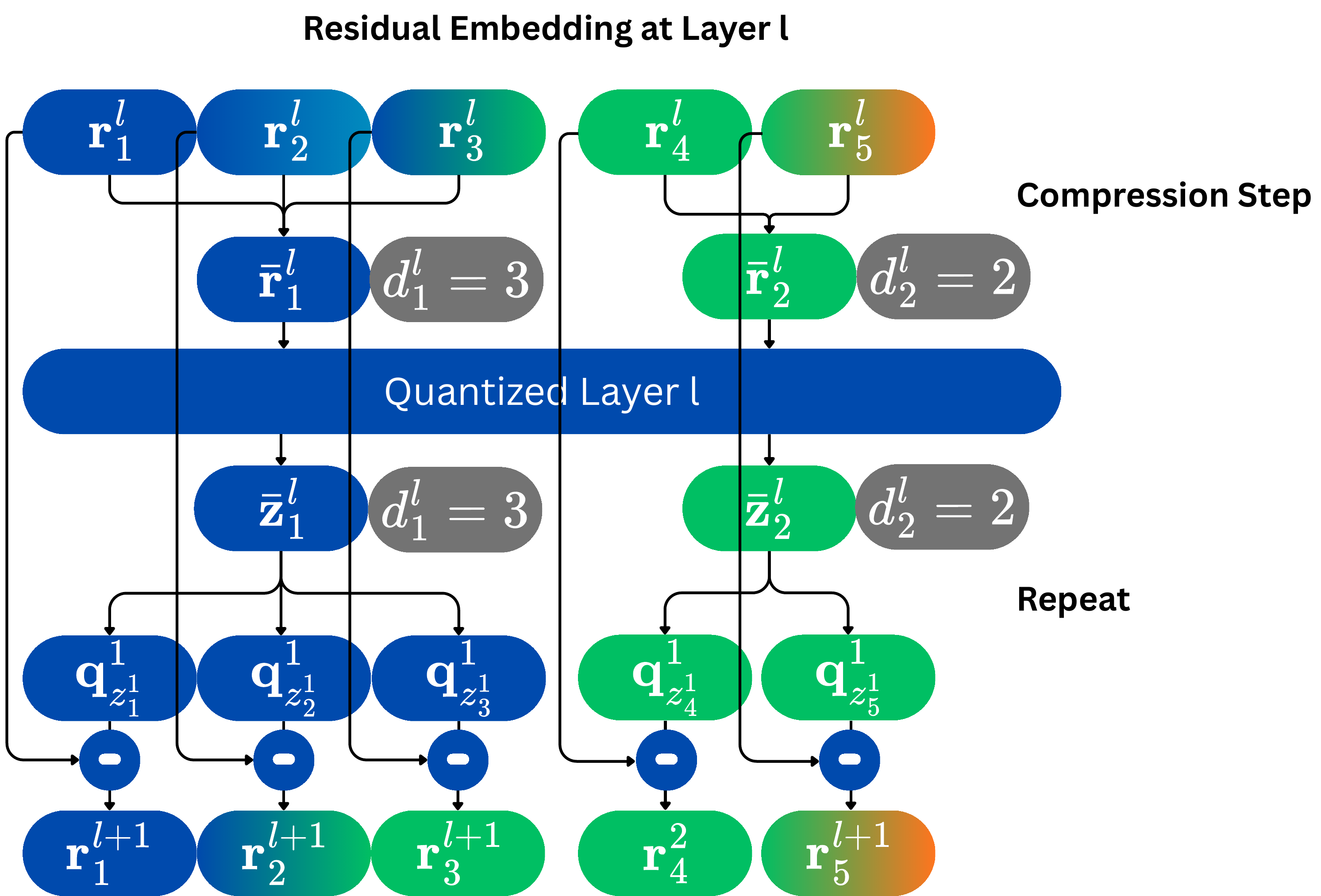}
    \caption{Overview of the LACE codec. At each quantization layer, the residual embeddings are independently compressed before being passed to the quantization layer. Frames with the same color belong to the same segment.}
    \label{fig:overview}
    \vspace{-15pt}
\end{figure}

Unlike prior dynamic frame rate codecs that apply a single compression step and share segmentation boundaries across quantization layers, LACE performs layer-wise compression on the residual embeddings. At layer $l$, we apply the compression step described in Section~\ref{sec:compression_background} to $\mathbf{R}^{l}$. This produces segment-level residual embeddings $\bar{\mathbf{R}}^{l} \in \mathbb{R}^{M^l \times D}$, durations $\mathbf{d}^{l} \in \mathbb{N}^{M^l}$, and layer-specific segmentation boundaries $\mathcal{B}^{l}$, where $M^l$ is the number of segments at layer $l$. The segment-level residual embeddings $\bar{\mathbf{R}}^{l}$ are then passed to quantization layer $l$, yielding segment-level codes $\bar{\mathbf{Z}}^{l} \in [1,V]^{M^l}$.

To compute the residual embeddings for the next layer, we expand $\bar{\mathbf{Z}}^{l}$ to frame-level codes $\hat{\mathbf{Z}}^l \in [1,V]^T$ by repeating each code according to $\mathbf{d}^{l}$. The next residual embedding is then computed as $\mathbf{r}_i^{l+1} = \mathbf{r}_i^{l} - \mathbf{q}^l_{\hat{z}^l_i}$. Unlike the standard RVQ update in Section~\ref{sec:preliminaries}, $\mathbf{q}^l_{\hat{z}^l_i}$ is obtained by quantizing a segment-level residual embedding, whereas $\mathbf{q}^l_{z^l_i}$ in the original update is obtained from a frame-level residual embedding. We use the frame-level residual embedding $\mathbf{r}_i^{l}$ on the right-hand side so that $\mathbf{r}_i^{l+1}$ captures both the compression error and the quantization error from layer $l$. Figure~\ref{fig:overview} illustrates layer-wise compression.

\subsection{Union Alignment} \label{sec:alignment}

\begin{figure}[t]
    \centering
    \includegraphics[width=.38\textwidth]{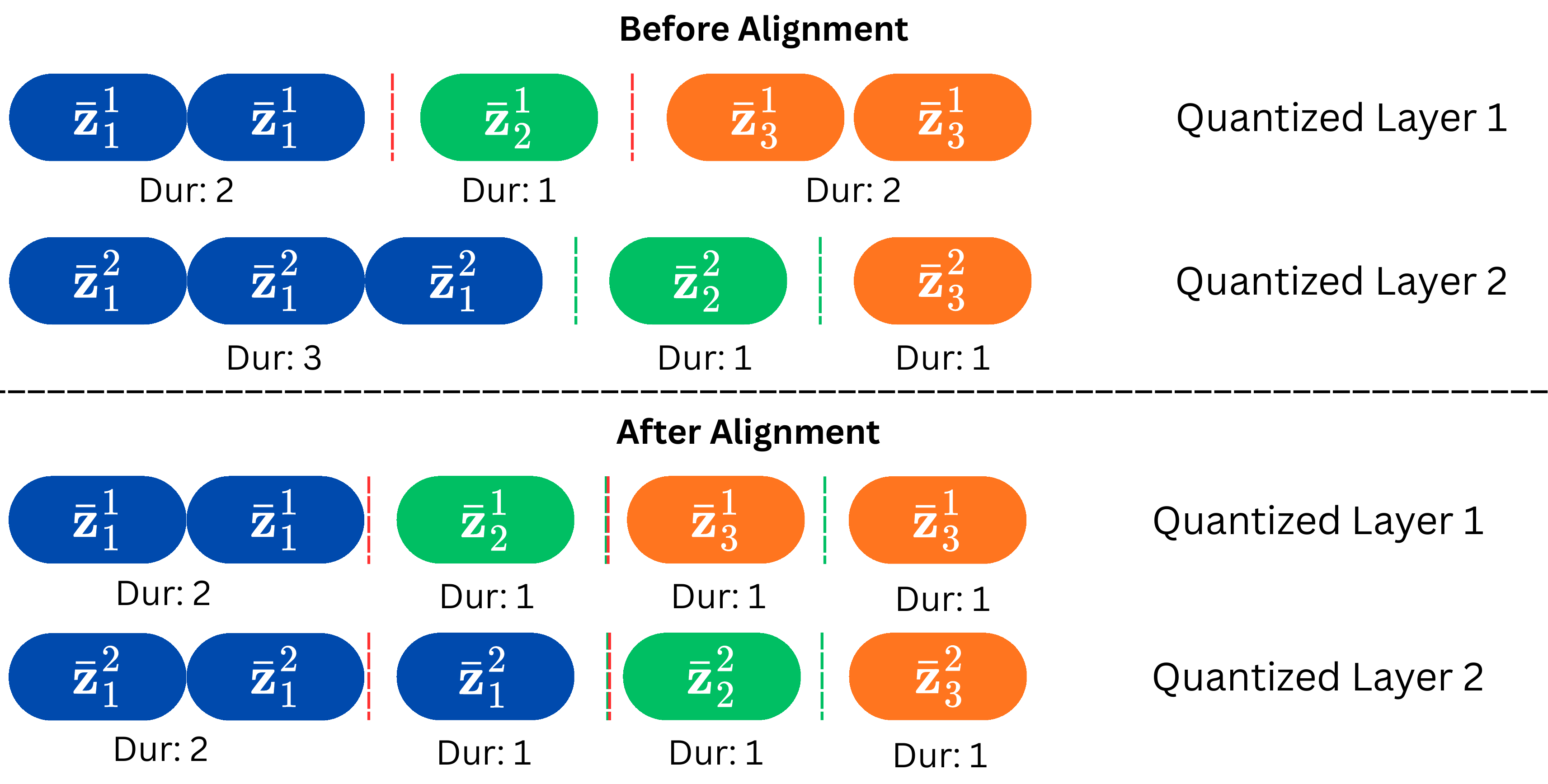}
    \caption{An overview of union alignment. \textit{Before alignment} (top), quantization layers 1 and 2 have different segmentation boundaries, resulting in inconsistent durations across layers. \textit{After alignment} (bottom), the union of segmentation boundaries from both layers is applied to all layers.
    }
    \label{fig:overview_union_alignment}
    \vspace{-15pt}
\end{figure}

The layer-wise compression produces different segmentation boundaries $\mathcal{B}^{l}$ for each quantization layer $l$. As a result, codes from different layers that cover the same time span can have different durations. This is problematic for TTS because the model would need to predict durations for each layer while ensuring that the total duration after upsampling is identical across layers. We address this with union alignment, which constructs shared segmentation boundaries by taking the union across layers: $\mathcal{B}^{\text{union}} = \bigcup_{l=1}^{C} \mathcal{B}^{l}$. We then re-segment the segment-level codes of every layer using $\mathcal{B}^{\text{union}}$. As shown in Figure~\ref{fig:overview_union_alignment}, when a boundary in $\mathcal{B}^{\text{union}}$ falls inside an existing segment, that segment is split into sub-segments. For example, the boundary $b_1=2$ from the first layer splits the first segment in the second layer. The resulting sub-segments keep the same code and receive new corresponding durations. After alignment, all layers share the same segmentation boundaries and durations, at the cost of increasing the effective frame rate.

\subsection{Boundary Anchor} \label{sec:anchor}

Union alignment can increase the effective frame rate because $|\mathcal{B}^{\text{union}}| \geq |\mathcal{B}^{l}|$ for all $l$. In the worst case, each layer introduces distinct boundaries and $|\mathcal{B}^{\text{union}}|$ approaches $T$, negating the compression benefit. To mitigate this, we introduce a boundary anchor. We choose an anchor layer $l^*$, where layers $l \leq l^*$ are allowed to introduce new boundaries, while deeper layers $l > l^*$ must reuse boundaries from earlier layers: $\mathcal{B}^{\text{anchor}} = \bigcup_{l=1}^{l^*} \mathcal{B}^{l}$. 
By restricting the compression step at layers $l > l^*$ to place boundaries only at positions in $\mathcal{B}^{\text{anchor}}$, we limit the growth of $|\mathcal{B}^{\text{union}}|$. Here, $l^*$ is a tunable hyperparameter that trades off between quality and the effective frame rate.

For DP-based compression, we implement this restriction with a constrained cost $\tilde{\ell}(j, s)$, which equals the segment loss $\ell(j, s)$ of Section~\ref{sec:compression_background} if both $j$ and $j{-}s$ lie in $\mathcal{B}^{\text{anchor}}$ and $+\infty$ otherwise. We solve the same recursion as Equation~\eqref{eq:dp_obj} with $\ell$ replaced by $\tilde{\ell}$, and use the unconstrained DP for layers $l \leq l^*$. While this formulation is specific to DP-based compression, the boundary anchor itself is model-agnostic. It can be applied to other compression models in Section~\ref{sec:compression_background} by restricting its candidate boundaries to $\mathcal{B}^{\text{anchor}}$.

\subsection{Theoretical Analysis}
\label{sec:proof}

Section~\ref{sec:anchor} introduced the anchor layer $l^*$ as a hyperparameter that trades off quality against the effective frame rate. We now analyze the quality side of this tradeoff: how the choice of $l^*$ affects the expected quantization error. In Sections~\ref{sec:compression_background} and~\ref{sec:multi_compression}, the compression step merges the frames before the quantization layer, and the codes are repeated back to the frame level after it. Since all repeated frames within a segment are identical, repeating before or after quantization yields the same result. We therefore combine merging and repetition into a single matrix multiplication $\mathbf{A}^l \mathbf{R}^l$, where $\mathbf{R}^l$ is the frame-level residual embedding at layer $l$. Concretely, $\mathbf{A}^l \in \mathbb{R}^{T\times T}$ is block-diagonal with one block per segment, and the block for a segment of length $d_m$ is the $d_m \times d_m$ matrix with every entry equal to $1/d_m$. Since $\mathbf{A}^l$ is symmetric and idempotent, the compression error $\mathbf{R}^l - \mathbf{A}^l \mathbf{R}^l$ is orthogonal to any frame-level embeddings that use the same segmentation boundaries as layer $l$. Let $\hat{\mathbf{Q}}^l \in \mathbb{R}^{T \times D}$ denote the frame-level quantized embeddings, which stack the selected codewords $\mathbf{q}^l_{\hat{z}^l_i}$, so the residual update of Section~\ref{sec:multi_compression} becomes $\mathbf{R}^{l+1} = \mathbf{R}^l - \hat{\mathbf{Q}}^l$.

The quantized embeddings $\hat{\mathbf{Q}}^l$ repeat one codeword across each segment. Subtracting them in the residual update can therefore only change the part of $\mathbf{R}^l$ that is constant within each segment. The compression error, which varies within segments, passes to the next layer untouched. When all layers share segmentation boundaries, the quantized embeddings of every layer are constant within the same segments, so the compression error survives all $C$ quantization layers. It becomes an error that quantization cannot remove. We formalize this intuition under three assumptions: (i) every layer merges at least two frames ($M^l < T$), so the compression error is strictly positive, (ii) each quantization layer is well-trained, approximately satisfying the centroid condition~\cite{gersho1992vector}, so that $\mathbb{E}[\|\mathbf{A}^l \mathbf{R}^l - \hat{\mathbf{Q}}^l\|_F^2] \leq \epsilon_l\, \mathbb{E}[\|\mathbf{A}^l \mathbf{R}^l\|_F^2]$ with $\epsilon_{\max} = \sup_l \epsilon_l < 1$, and (iii) the fraction of expected residual energy captured by the compression step, $\alpha_l = \mathbb{E}[\|\mathbf{A}^l \mathbf{R}^l\|_F^2] / \mathbb{E}[\|\mathbf{R}^l\|_F^2]$, is bounded below by $\alpha_{\min} = \inf_l \alpha_l > 0$.

Under these assumptions, the expected residual error after all $C$ quantization layers with anchor layer $l^*$ satisfies\footnote{The analysis assumes that layers $l > l^*$ reuse the segmentation boundaries of layer $l^*$ exactly, a special case of the boundary anchor in Section~\ref{sec:anchor}.}
\begin{align}
    \mathbb{E}[\|\mathbf{R}^{C+1}\|_F^2]
    \leq \Big[ 1 - \alpha_{l^*} \big( 1 - \epsilon_{\max}^{\,C - l^* + 1} \big) \Big]
    \prod_{l=1}^{l^*-1} \lambda_l\,
    \mathbb{E}[\|\mathbf{H}\|_F^2],
    \label{eq:main_bound}
\end{align}
where $\mathbf{R}^{C+1}$ is the residual after the last layer and $\lambda_l = 1 - \alpha_l (1 - \epsilon_l) < 1$. As $C \to \infty$, Equation~\eqref{eq:main_bound} reveals a hierarchy. A single compression, which shares the same segmentation boundaries across layers ($l^*{=}1$), converges to a constant error floor $(1-\alpha_1)\,\mathbb{E}[\|\mathbf{H}\|_F^2]$, which is exactly the compression error of the first layer. The boundary anchor reduces this floor by the factor $\prod_{l=1}^{l^*-1} \lambda_l < 1$. Full layer-wise compression ($l^*{=}C$) drives the bound to zero. Therefore, recomputing boundaries at each quantization layer removes the error floor inherent to shared segmentation boundaries. The full derivation is provided in the supplementary material. This bound concerns the codec quantization error only, not downstream TTS, where union alignment (Section~\ref{sec:alignment}) also changes the token sequence.

\subsection{TTS Training with LACE Tokens}
\label{sec:tts_model}
Unlike standard TTS models \cite{chen2025valle, musicgen}, a TTS model trained on LACE tokens must also predict the durations $\mathbf{d}$. After union alignment, all $C$ quantization layers share the segmentation boundaries $\mathcal{B}^{\text{union}}$ and durations $\mathbf{d}$. Therefore, the model predicts durations only for the first quantization layer and applies the predicted durations to all subsequent layers.

We use an autoregressive decoder-only Transformer that predicts the segment-level codes $\bar{\mathbf{Z}}$ in a delay-pattern format~\cite{musicgen}. At each generation step for the first-layer code, the model additionally predicts a duration for that code. Duration prediction is formulated as a classification task. We add a duration head alongside the code prediction head and include a learned duration embedding in the input representation, which is added to the code embedding. The model is trained with cross-entropy loss for code prediction and focal loss for duration prediction to address class imbalance, with loss weights of 1 and 3, respectively. At inference, we use nucleus sampling with a top-$p$ value of 0.8 and a temperature of 1.0.

\begin{table*}[t]
\centering
\caption{Reconstruction performance across different codecs and compression models on LibriTTS \texttt{test-clean}. 
All baseline results use a single compression step without LACE. ``+ LACE'' denotes our proposed layer-wise compression. The frame rate column reports the effective frame rate and codec frame rate for codecs with and without compression models, respectively. 
For threshold-based compression models, we report the average effective frame rate across layers. Baseline rows are re-implemented using the same backbones. The results reported by the cited systems are not directly comparable.
}
\vspace{-5pt}

\label{tab:reconstruct}

\renewcommand{\arraystretch}{0.92}
\begin{tabular}{ll|cc|ccccc}
\toprule
\textbf{Codec} &
\textbf{Compression} &
\textbf{Frame Rate} &
\textbf{Bitrate} &
\textbf{WER} $\downarrow$ &
\textbf{UTMOS} $\uparrow$ &
\textbf{PESQ} $\uparrow$ &
\textbf{STOI} $\uparrow$ &
\textbf{SpkSim} $\uparrow$ \\
&
&
\textbf{(Hz)} &
\textbf{(kbps)} &
(\%) & & & &  \\
\midrule
Ground Truth & --                          & - & - & 2.02 & 4.06 & - & - & - \\
\midrule
DAC~\cite{dac_codec}         & --                          & 75 & 24.00 & 2.16 & 3.95 & 3.59 & 0.97 & 0.99 \\
Encodec~\cite{encodec_codec}      & --                          & 75 & 24.00 & 2.09 & 3.93 & 3.25 & 0.96 & 0.98  \\
SoundStream~\cite{soundstream_codec}  & --                          & 75 & 24.00 & 2.41 & 3.61 & 2.56 & 0.93 & 0.97  \\

\midrule

\multirow{6}{*}{Finetuned DAC}
& DP~\cite{codecslime}         & 27 & 8.69 & 4.92 & 2.49 & 1.61 & 0.86 & 0.94 \\
& DP + LACE            & 22.5 & 8.64 & \textbf{2.22} & \textbf{3.81} & \textbf{3.08} & \textbf{0.95} & \textbf{0.98} \\
\cmidrule{2-9}
& Cosine~\cite{li2025flexicodec}        & 27 & 8.69  & 6.90 & 1.84 & 1.29 & 0.81 & 0.90 \\
& Cosine + LACE            & 22.5 & 8.64 & \textbf{2.57} & \textbf{3.55} & \textbf{2.73} & \textbf{0.94} & \textbf{0.97} \\
\cmidrule{2-9}
& Density Clustering~\cite{zheng2025sayless}  & 27 & 8.69 & 4.25 & 2.19 & 1.45 & 0.85 & 0.94 \\
& Density Clustering + LACE  & 22.5 & 8.64  & \textbf{2.62} & \textbf{2.89} & \textbf{2.08} & \textbf{0.95} & \textbf{0.96} \\

\midrule

\multirow{2}{*}{Finetuned SoundStream}
& DP & 27 & 8.69 & 7.37 & 2.54 & 1.52 & 0.83 & 0.92  \\
& DP + LACE  & 22.5 & 8.64 & \textbf{2.78} & \textbf{3.60} & \textbf{2.45} & \textbf{0.92} & \textbf{0.97} \\

\midrule

\multirow{2}{*}{Finetuned Encodec}
& DP & 27 & 8.69 & 3.10 & 3.08 & 1.99 & 0.89 & 0.96  \\
& DP + LACE  & 22.5 & 8.64 & \textbf{2.18} & \textbf{3.83} & \textbf{2.86} & \textbf{0.95} & \textbf{0.98} \\

\bottomrule
\end{tabular}
\vspace{-15pt}
\end{table*}

\section{Experimental Setting}

\subsection{Dataset}
We use the LibriTTS dataset~\cite{zen19_interspeech} at 24 kHz sampling rate. The training set consists of \texttt{train-clean-100}, \texttt{train-clean-360}, and \texttt{train-other-500}. Evaluation is performed on \texttt{test-clean}. The same dataset is used for both reconstruction and TTS experiments. For TTS, a reference utterance is randomly selected from a different utterance of the same speaker to condition the model on speaker identity.

\subsection{Model Architecture}
\subsubsection{Codec Models}
We build our method on three neural audio codecs: SoundStream \cite{soundstream_codec}, EnCodec \cite{encodec_codec}, and the Descript Audio Codec (DAC) \cite{dac_codec}, all pretrained on LibriTTS using ESPnet-Codec \cite{shi2024espnetcodec} recipes\footnote{\url{https://huggingface.co/espnet/libritts_soundstream24k}, \url{https://huggingface.co/espnet/libritts_encodec_24k}, \url{https://huggingface.co/espnet/libritts_dac_24k}}. We integrate the proposed layer-wise compression into the quantization layers (Section~\ref{sec:multi_compression}) and fine-tune each model end-to-end from the pretrained weights for 80k iterations using the original codec objective. We use Adam~\cite{kingma2015adam} with learning rate $10^{-4}$, an exponential decay schedule (rate $0.9998$ per step, minimum $10^{-5}$), and shared optimization settings for the generator and discriminator. All experiments use 4 V100 32GB GPUs.

\subsubsection{TTS Model}
The TTS model is a 24-layer Transformer with 16 attention heads, model dimension 1024, feedforward dimension 4096, and dropout 0.1. Input transcripts are converted to phoneme sequences using the espeak-ng phonemizer~\cite{Bernard2021}. A randomly truncated 3-second reference utterance is prepended to condition the model on speaker identity. The model is trained for 40k iterations on 4 V100 32GB GPUs using Adam with learning rate $10^{-4}$ and 4000 warmup steps.

\subsection{Baselines}
We compare LACE with CodecSlime~\cite{codecslime}, FlexiCodec~\cite{li2025flexicodec}, and VARSTok~\cite{zheng2025sayless}. Because these methods use different codec backbones and training setups, we re-implement them on LibriTTS with the same codec backbones and fine-tuning configuration, changing only the compression model. These baselines use single compression, where the compression step is applied once before the quantization layers so all layers share the same segmentation boundaries. In contrast, LACE applies compression independently at each quantization layer.

\subsection{Rate Computation}
\label{sec:rate_computation}
For reconstruction, we report the effective frame rate before union alignment, since alignment is not required. For threshold-based compression models that yield different effective frame rates across layers, we report the mean effective frame rate across layers. The bitrate accounts for both code and duration bits, where each duration costs $\log_2 U$ bits per segment. For a fair comparison, we compare at matched bitrate ($\sim 8.6$ kbps). Note that because LACE adds one duration sequence per layer, it reaches this bitrate at a lower effective frame rate than single compression (22.5 vs. 27 Hz).

\subsection{Evaluation Metrics}

\paragraph{Reconstruction.}
We compute word error rate (WER) with Whisper-Large~\cite{whisper} by transcribing the reconstructed audio. UTMOS~\cite{utmos} assesses perceptual naturalness, while PESQ~\cite{pesq} and STOI~\cite{stoi} measure signal quality and intelligibility. For speaker similarity (SpkSim), we compute the cosine similarity between X-vectors from the generated and reference audio, extracted with WavLM-base~\cite{chen2022wavlm} fine-tuned for speaker verification\footnote{\url{https://huggingface.co/microsoft/wavlm-base-plus-sv}}. All metrics use VERSA~\cite{shi2025versa} to compute.

\paragraph{TTS.}
Since TTS generation is non-deterministic, we generate 10 utterances per input and select the one with the lowest WER under Whisper-Small, following prior work~\cite{mousavi2025discrete}. The selected utterance is evaluated with Whisper-Large for WER, along with UTMOS and SpkSim. We also report the real-time factor (RTF), the ratio of generation time to audio duration, to assess inference efficiency. To complement these objective metrics, we conduct a human evaluation using mean opinion score (MOS) and speaker similarity mean opinion score (SMOS) to evaluate naturalness and speaker similarity, respectively. Specifically, we randomly sample 25 test utterances and synthesize them with each TTS system. Then, we collect ratings from 35 evaluators.

\section{Results}
\subsection{Reconstruction Task}

\begin{figure}[t]
    \centering
    \includegraphics[width=.80\linewidth]{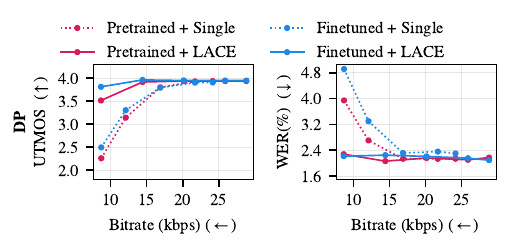}
    \vspace{-17pt}
    \caption{Rate-distortion comparison between LACE and Single compression on LibriTTS \texttt{test-clean} using DAC as the codec backbone and DP compression. 
    }
    \label{fig:rate_distortion}
    \vspace{-8pt}
\end{figure}

We evaluate reconstruction quality across three codec backbones (DAC~\cite{dac_codec}, SoundStream~\cite{soundstream_codec}, EnCodec~\cite{encodec_codec}) and three compression models (DP~\cite{codecslime}, Cosine Similarity~\cite{li2025flexicodec}, and Density Clustering~\cite{zheng2025sayless}). All systems are matched at a similar bitrate ($\sim8.6$ kbps), resulting in different effective frame rates. Table~\ref{tab:reconstruct} shows that LACE surpasses the single-compression baselines in every configuration, indicating that its benefit is agnostic to both the compression model and the codec backbone. Figure~\ref{fig:rate_distortion} provides a finer-grained view through the rate-quality curve. LACE outperforms single compression at every bitrates, even on the pretrained model without fine-tuning. This indicates that the improvement comes from LACE itself, while fine-tuning provides further gains.

\subsection{Text-to-Speech Task}
\label{sec:tts_results}
\begin{table}[t]
\centering
\setlength{\tabcolsep}{1.5pt}
\caption{TTS evaluation on LibriTTS \texttt{test-clean}. 
Rate is the effective frame rate after union alignment. We report MOS and SMOS, with 95\% confidence intervals.
RTF is measured on a single V100 GPU for one candidate.}
\vspace{-5pt}
\label{tab:tts_results}
\resizebox{\linewidth}{!}{
\begin{tabular}{l|ccc|ccc|cc|c}
\toprule
\textbf{Method} & $\gamma$ & $l^*$ & \textbf{Rate} & \textbf{WER} $\downarrow$ & \textbf{UTMOS} $\uparrow$ & \textbf{SpkSim} $\uparrow$ & \textbf{MOS} $\uparrow$ & \textbf{SMOS} $\uparrow$ & \textbf{RTF} $\downarrow$ \\
 & & & (Hz) & (\%) & & & & & \\
\midrule
No compression & -- & -- & 75.0 & 2.31 & 4.18 & 0.91 & 3.94  $\pm$ 0.05 & 3.94 $\pm$ 0.06 & 1.95 \\
\midrule
Single & 0.5 & -- & 37.5 & 30.73 & 1.79 & 0.84 & 3.80 $\pm$ 0.06 & 3.80 $\pm$ 0.07 & 1.01 \\
LACE & 0.5 & 2 & 52.5 & \textbf{4.93} & 2.69 & 0.86 & 3.88 $\pm$ 0.06 & \textbf{3.86 $\pm$ 0.07} & 1.58 \\
LACE & 0.5 & 3 & 58.0 & 8.52 & \textbf{2.70} & \textbf{0.88} & \textbf{3.89 $\pm$ 0.06} & 3.84 $\pm$ 0.06 &  1.68\\
\midrule
Single & 0.7 & -- & 52.5 & 6.77 & 2.59 & 0.85 & 3.85 $\pm$ 0.06 & 3.82 $\pm$ 0.07 & 1.45 \\
LACE & 0.7 & 2 & 61.8 & \textbf{5.11} & \textbf{3.45} & \textbf{0.90} & \textbf{3.96 $\pm$ 0.05} & \textbf{3.90 $\pm$ 0.06} & 1.73 \\
\bottomrule
\end{tabular}}
\vspace{-10pt}
\end{table}

We train and evaluate the TTS model using DAC as the codec backbone with DP compression. We compare against two baselines: DAC without compression and DAC with single DP compression.
Table~\ref{tab:tts_results} shows that, at the same target compression rate $\gamma$, LACE outperforms single compression on all quality metrics. This improvement comes at the cost of a higher RTF because union alignment increases the effective frame rate. Although the MOS scores of single compression with $\gamma=0.5$ and no compression are close, the paired $t$-test shows that raters can still distinguish between them $(p<0.01)$.
Compared to the no-compression baseline, LACE achieves a lower RTF by reducing the effective frame rate. This comes at a quality cost, partly because the compressed codec has lower reconstruction quality (Table~\ref{tab:reconstruct}). 
 The target compression rate $\gamma$ directly controls the quality--efficiency tradeoff. Increasing $\gamma$ improves synthesis quality while increasing RTF. In contrast, increasing the anchor layer $l^*$ beyond 2 does not improve quality and degrades WER. We hypothesize that this behavior arises from the union alignment step. A larger $l^*$ causes long segments to be split into shorter ones, introducing repeated discrete code IDs after alignment. These repeated code IDs bias the TTS model toward repeatedly predicting the same code IDs during generation. Incorporating repetition-aware sampling~\cite{chen2025valle} to prevent repetitive code prediction is a promising direction for future work. To ensure that the observed WER trends are robust to the reranking, we computed the mean WER over 10 generated candidates using Whisper-Small. The ranking remains consistent. LACE achieves 14\% WER compared with 61\% for single compression at $\gamma=0.5$, and 14\% compared with 21\% at $\gamma=0.7$.

\begin{figure}[t]
            \centering
            \includegraphics[width=.78\linewidth]{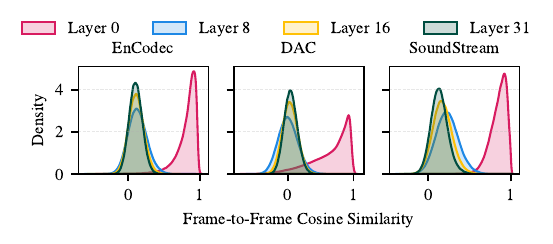}
            \vspace{-15pt}
            \caption{Distribution of cosine similarity between consecutive frame-level residual embeddings across quantization layers. 
            }
            \label{fig:cosine_sim}
            \vspace{-15pt}
\end{figure}

\begin{figure}[t]
    \centering
    \includegraphics[width=.78\linewidth]{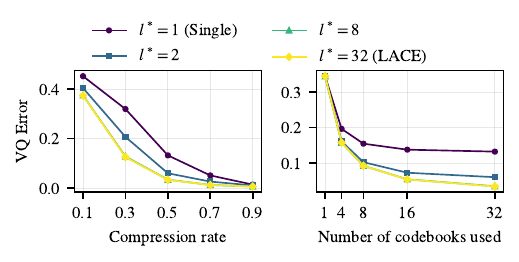}
    \vspace{-15pt}
    \caption{Empirical validation of the theoretical analysis using the pretrained DAC model with DP compression. \textit{Left}: quantization error at various compression rates $\gamma$ using all 32 quantization layers. \textit{Right}: quantization error as a function of the number of quantization layers, at $\gamma{=}0.5$. 
    }
    \label{fig:vq_error}
    \vspace{-20pt}
\end{figure}

\subsection{Analysis}
\label{sec:analysis}
We perform two analyses to better understand LACE. The first examines how fast the residual embeddings change at each quantization layer. The second empirically validates the quantization error bound from Section~\ref{sec:proof}.

\subsubsection{Histogram of Cosine Similarity}\label{sec:analysis_cosine}On the pretrained codecs, we compute the cosine similarity between consecutive frame-level residual embeddings, $\sigma^l_t = \cos(\mathbf{r}^l_t, \mathbf{r}^l_{t+1})$. Figure~\ref{fig:cosine_sim} shows that its distribution concentrates around $1$ at earlier layers and gradually shifts toward $0$ at deeper layers. This indicates that residual embeddings change more rapidly at deeper layers and confirms that different layers change at different rates, motivating layer-wise compression. Moreover, the slower variation of earlier layers suggests that segmentation decisions have a larger impact on these layers than on deeper layers. We therefore let earlier layers choose its own segmentation boundaries before deeper layers, justifying the boundary anchor (Section~\ref{sec:anchor}).

\subsubsection{Empirical Result of VQ Error}
While Figure~\ref{fig:cosine_sim} only motivates layer-wise compression, this analysis directly measures the error caused by shared segmentation boundaries. We measure the quantization error $\mathbb{E}[\|\mathbf{R}^{C+1}\|_F^2]$ on the pretrained DAC model with DP compression while varying the anchor layer $l^*$. Figure~\ref{fig:vq_error} shows that LACE yields consistently lower quantization error than single compression, matching Equation~\eqref{eq:main_bound}. The right panel shows the gap widening as more layers are used. Single compression converges to a constant error floor, while LACE keeps reducing the error. The left panel shows the role of $\gamma$. As $\gamma \to 1$, compression merges almost no frames, so $\alpha_1 \to 1$ and the floor $(1-\alpha_1)\,\mathbb{E}[\|\mathbf{H}\|_F^2]$ vanishes, explaining the narrow gap at high compression rates. At lower rates the floor grows and the benefit of LACE becomes more pronounced.



\section{Conclusion}
We presented LACE, a dynamic frame rate codec that applies an independent compression step at each quantization layer. Unlike prior methods that force all layers to share the same segmentation boundaries, LACE allows each layer to choose its own segmentation boundaries. To enable downstream TTS training, we introduced union alignment, which constructs shared segmentation boundaries across layers, and boundary anchor, which limits the growth of the effective frame rate. Experiments on LibriTTS show that LACE consistently outperforms single-compression baselines across multiple codec backbones and compression models on the reconstruction task. Applied to TTS, LACE improves inference efficiency while maintaining competitive quality.

\section*{Acknowledgment}
Experiments of this work used the Bridges2 system at PSC and Delta and DeltaAI system at NCSA through allocations CIS210014 and IRI120008P from the Advanced Cyberinfrastructure Coordination Ecosystem: Services \& Support (ACCESS) program, supported by National Science Foundation grants \#2138259, \#2138286, \#2138307, \#2137603, and \#2138296.

\section*{Generative AI Use Disclosure}
Generative AI tools were used only to edit and polish the manuscript language and to assist with code writing. All research ideas, experimental design, implementation, analysis, and reported results are the work of the authors. The conceptual framing, methodology, and scientific contributions are entirely human-generated.

\bibliographystyle{IEEEtran}
\bibliography{mybib}

\end{document}


\section{Supplementary Material}

All sections, equation, and figure references below refer to the main paper unless stated otherwise.

\subsection{Upper Bound of the Expected Quantization Error}
\label{sec:proof_full}

\paragraph{Setup.}
Let $\mathbf{H} \in \mathbb{R}^{T \times D}$ be the frame-level embeddings drawn from an in-domain speech distribution. At layer $l$, the compression step performs two operations: (1) averaging the frame-level embeddings within each segment, and (2) repeating each segment-level embedding according to its duration. In Sections II-B and III-A, the repetition happens after the quantization layer rather than before, since all repeated frames within a segment are identical, the two orders yield the same result. We therefore combine both operations into a single matrix multiplication $\mathbf{A}^l \mathbf{R}^l$, where $\mathbf{A}^l \in \mathbb{R}^{T \times T}$ is the compression matrix at layer $l$ and $\mathbf{R}^l$ is the frame-level residual embedding from Section III-A, with $\mathbf{R}^1 = \mathbf{H}$. The matrix $\mathbf{A}^l$ is block-diagonal: the block of the $m$-th segment has every entry equal to $1/d_m$, where $d_m$ is the segment duration. From this structure, $\mathbf{A}^l$ is symmetric ($(\mathbf{A}^l)^\top = \mathbf{A}^l$) and idempotent ($(\mathbf{A}^l)^2 = \mathbf{A}^l$).

Let $\text{VQ}_l(\cdot)$ denote quantization at layer $l$ followed by upsampling, so that the frame-level quantized embedding is $\hat{\mathbf{Q}}^l = \text{VQ}_l(\mathbf{A}^l \mathbf{R}^l)$ and the residual update of Section III-A becomes $\mathbf{R}^{l+1} = \mathbf{R}^l - \hat{\mathbf{Q}}^l$. Since $\hat{\mathbf{Q}}^l$ is constant within each segment of layer $l$, applying the compression matrix leaves it unchanged:
\begin{align}
    \mathbf{A}^l \hat{\mathbf{Q}}^l = \hat{\mathbf{Q}}^l. \label{eq:code_in_range}
\end{align}

We analyze the boundary anchor setting of Section III-C: layers $l \leq l^*$ compute new boundaries, while layers $l > l^*$ reuse the compression matrix of layer $l^*$, denoted $\mathbf{A}^\star = \mathbf{A}^{l^*}$.

\paragraph{Assumptions.}
\begin{enumerate}
    \item \emph{Strict compression:} every layer merges at least two frames ($M^l < T$), so the compression error is strictly positive: $\mathbb{E}[\|\mathbf{R}^l - \mathbf{A}^l \mathbf{R}^l\|_F^2] > 0$.
    \item \emph{Centroid condition:} each quantization layer is well-trained. Its codewords satisfy the centroid condition~\cite{gersho1992vector}, i.e., each codeword equals the conditional mean of the inputs assigned to it. We further assume that each layer captures positive energy, $\mathbb{E}[\|\hat{\mathbf{Q}}^l\|_F^2] > 0$, and the resulting distortion ratios (Lemma~2) are uniformly bounded by $\epsilon_{\max} = \sup_l \epsilon_l < 1$.
    \item \emph{Positive energy capture:} we define
    \begin{align}
        \alpha_l = \frac{\mathbb{E}[\|\mathbf{A}^l \mathbf{R}^l\|_F^2]}{\mathbb{E}[\|\mathbf{R}^l\|_F^2]}
        \label{eq:energy_ratio}
    \end{align}
    as the fraction of expected residual energy captured by the compression step at layer $l$, and assume it is uniformly bounded below by $\alpha_{\min} = \inf_l \alpha_l > 0$. Lemma~1 shows $\alpha_l \leq 1$, together with Assumption~1 and Equation~\eqref{eq:pythagoras}, $\alpha_l < 1$.
\end{enumerate}

\noindent\textbf{Lemma 1 (Compression never increases energy).}
For any $\mathbf{R} \in \mathbb{R}^{T \times D}$ and any compression matrix $\mathbf{A}$,
\begin{align}
    \|\mathbf{A}\mathbf{R}\|_F^2 \leq \|\mathbf{R}\|_F^2.
\end{align}
\noindent\textit{Proof.} The compression matrix acts independently on each segment and each feature dimension, so it suffices to prove the claim for a single segment and a single dimension. Let $x_1, \dots, x_{d_m}$ denote the scalar values within a segment of duration $d_m$. Compression replaces every value with the segment mean $\mu = \frac{1}{d_m} \sum_{i=1}^{d_m} x_i$, so the energy of the segment changes from $\sum_{i=1}^{d_m} x_i^2$ to $d_m \mu^2 = \frac{1}{d_m} \big( \sum_{i=1}^{d_m} x_i \big)^2$. By the Cauchy--Schwarz inequality with the all-ones vector,
\begin{align}
    \Big( \sum_{i=1}^{d_m} x_i \cdot 1 \Big)^2
    \leq \Big( \sum_{i=1}^{d_m} x_i^2 \Big) \Big( \sum_{i=1}^{d_m} 1^2 \Big)
    = d_m \sum_{i=1}^{d_m} x_i^2,
\end{align}
which gives $d_m \mu^2 \leq \sum_{i=1}^{d_m} x_i^2$. Summing over all segments and all feature dimensions yields the claim. \hfill$\blacksquare$

\noindent\textbf{Lemma 2 (Expected rate-distortion bound).}
Under Assumption~2, the expected quantization error at layer $l$ satisfies
\begin{align}
    \mathbb{E}\big[\|\mathbf{A}^l \mathbf{R}^l - \hat{\mathbf{Q}}^l\|_F^2\big] \leq \epsilon_l\, \mathbb{E}\big[\|\mathbf{A}^l \mathbf{R}^l\|_F^2\big],
    \label{eq:rd_bound}
\end{align}
where $\epsilon_l = 1 - \mathbb{E}[\|\hat{\mathbf{Q}}^l\|_F^2] / \mathbb{E}[\|\mathbf{A}^l \mathbf{R}^l\|_F^2] < 1$.

\noindent\textit{Proof.} Write $\mathbf{X} = \mathbf{A}^l \mathbf{R}^l$ for the quantization input and $\hat{\mathbf{X}} = \hat{\mathbf{Q}}^l$ for its output. The quantization partitions the input space into regions $\{\mathcal{V}_k\}$ and maps every input in region $\mathcal{V}_k$ to the output $\mathbf{c}_k$. By the centroid condition, $\mathbf{c}_k = \mathbb{E}[\mathbf{X} \mid \mathbf{X} \in \mathcal{V}_k]$, so the expected quantization error within each region is zero:
\begin{align}
    \mathbb{E}[\mathbf{X} - \mathbf{c}_k \mid \mathbf{X} \in \mathcal{V}_k]
    = \mathbb{E}[\mathbf{X} \mid \mathbf{X} \in \mathcal{V}_k] - \mathbf{c}_k = \mathbf{0}.
\end{align}
By the law of total expectation, the quantization error is therefore orthogonal to the quantization output in expectation:
\begin{align}
    \mathbb{E}\big[\langle \mathbf{X} - \hat{\mathbf{X}},\, \hat{\mathbf{X}} \rangle_F\big]
    &= \sum_k \mathbb{P}(\mathbf{X} \in \mathcal{V}_k)\, \big\langle \mathbb{E}[\mathbf{X} - \mathbf{c}_k \mid \mathbf{X} \in \mathcal{V}_k],\, \mathbf{c}_k \big\rangle_F \nonumber \\
    &= 0,
\end{align}
where $\mathbf{c}_k$ is pulled out of the conditional expectation because it is constant within its region. The cross-term thus vanishes in the expected Pythagorean decomposition:
\begin{align}
    \mathbb{E}[\|\mathbf{X}\|_F^2] = \mathbb{E}[\|\hat{\mathbf{X}}\|_F^2] + \mathbb{E}[\|\mathbf{X} - \hat{\mathbf{X}}\|_F^2].
\end{align}
Rearranging gives the exact equality
\begin{align}
    \mathbb{E}[\|\mathbf{X} - \hat{\mathbf{X}}\|_F^2]
    = \Big( 1 - \frac{\mathbb{E}[\|\hat{\mathbf{X}}\|_F^2]}{\mathbb{E}[\|\mathbf{X}\|_F^2]} \Big) \mathbb{E}[\|\mathbf{X}\|_F^2]
    = \epsilon_l\, \mathbb{E}[\|\mathbf{X}\|_F^2],
\end{align}
and Assumption~2 ($\mathbb{E}[\|\hat{\mathbf{X}}\|_F^2] > 0$) gives $\epsilon_l < 1$. \hfill$\blacksquare$

\noindent\textbf{Lemma 3 (Orthogonality).}
For any $\mathbf{R}, \mathbf{S} \in \mathbb{R}^{T \times D}$ and any compression matrix $\mathbf{A}$, the compression error $\mathbf{R} - \mathbf{A}\mathbf{R}$ is orthogonal to any compressed embedding $\mathbf{A}\mathbf{S}$ under the Frobenius inner product:
\begin{align}
    \langle \mathbf{R} - \mathbf{A}\mathbf{R},\, \mathbf{A}\mathbf{S} \rangle_F
    = \mathrm{Tr}\big(\mathbf{R}^\top (\mathbf{A} - \mathbf{A}^2) \mathbf{S}\big) = 0,
\end{align}
using symmetry and idempotence of $\mathbf{A}$. In particular, choosing $\mathbf{S} = \mathbf{R}$ yields the Pythagorean identity:
\begin{align}
    \|\mathbf{R}\|_F^2 = \|\mathbf{A}\mathbf{R}\|_F^2 + \|\mathbf{R} - \mathbf{A}\mathbf{R}\|_F^2.
    \label{eq:pythagoras}
\end{align}

\noindent\textbf{Lemma 4 (Per-layer energy contraction).}
For every layer $l$ that computes new boundaries,
\begin{align}
    \mathbb{E}[\|\mathbf{R}^{l+1}\|_F^2] \leq \lambda_l\, \mathbb{E}[\|\mathbf{R}^{l}\|_F^2],
\end{align}
where $\lambda_l = 1 - \alpha_l (1 - \epsilon_l) < 1$.

\noindent\textit{Proof.} Decompose the residual update as $\mathbf{R}^{l+1} = (\mathbf{R}^l - \mathbf{A}^l \mathbf{R}^l) + (\mathbf{A}^l \mathbf{R}^l - \hat{\mathbf{Q}}^l)$. By Equation~\eqref{eq:code_in_range}, the second term can be rewritten as a compressed embedding:
\begin{align}
    \mathbf{A}^l \mathbf{R}^l - \hat{\mathbf{Q}}^l
    = \mathbf{A}^l \mathbf{R}^l - \mathbf{A}^l \hat{\mathbf{Q}}^l
    = \mathbf{A}^l (\mathbf{R}^l - \hat{\mathbf{Q}}^l).
\end{align}
Therefore, the two terms are orthogonal by Lemma~3 with $\mathbf{S} = \mathbf{R}^l - \hat{\mathbf{Q}}^l$, and their squared norms add:
\begin{align}
    \mathbb{E}[\|\mathbf{R}^{l+1}\|_F^2]
    &= \mathbb{E}[\|\mathbf{R}^l - \mathbf{A}^l \mathbf{R}^l\|_F^2] + \mathbb{E}[\|\mathbf{A}^l \mathbf{R}^l - \hat{\mathbf{Q}}^l\|_F^2] \nonumber \\
    &\leq (1 - \alpha_l)\, \mathbb{E}[\|\mathbf{R}^l\|_F^2] + \epsilon_l\, \alpha_l\, \mathbb{E}[\|\mathbf{R}^l\|_F^2] \nonumber \\
    &= \lambda_l\, \mathbb{E}[\|\mathbf{R}^l\|_F^2].
\end{align}
The first term follows from Equation~\eqref{eq:pythagoras} and the definition of $\alpha_l$ in Equation~\eqref{eq:energy_ratio}, which together give $\mathbb{E}[\|\mathbf{R}^l - \mathbf{A}^l \mathbf{R}^l\|_F^2] = (1 - \alpha_l)\, \mathbb{E}[\|\mathbf{R}^l\|_F^2]$. The second term applies Equation~\eqref{eq:rd_bound} and then Equation~\eqref{eq:energy_ratio}: $\mathbb{E}[\|\mathbf{A}^l \mathbf{R}^l - \hat{\mathbf{Q}}^l\|_F^2] \leq \epsilon_l\, \mathbb{E}[\|\mathbf{A}^l \mathbf{R}^l\|_F^2] = \epsilon_l\, \alpha_l\, \mathbb{E}[\|\mathbf{R}^l\|_F^2]$. \hfill$\blacksquare$

\noindent\textbf{Lemma 5 (Invariance of the compression error).}
If $\mathbf{A}^l = \mathbf{A}^\star$ for all $l \geq l^*$, then for all such layers:
\begin{align}
    \mathbf{R}^{l+1} - \mathbf{A}^\star \mathbf{R}^{l+1} = \mathbf{R}^{l} - \mathbf{A}^\star \mathbf{R}^{l}.
\end{align}
\noindent\textit{Proof.} Applying $\mathbf{A}^\star$ to the residual update $\mathbf{R}^{l+1} = \mathbf{R}^l - \hat{\mathbf{Q}}^l$ and using Equation~\eqref{eq:code_in_range}:
\begin{align}
    \mathbf{A}^\star \mathbf{R}^{l+1} = \mathbf{A}^\star \mathbf{R}^{l} - \hat{\mathbf{Q}}^l.
    \label{eq:insubspace_recursion}
\end{align}
Subtracting Equation~\eqref{eq:insubspace_recursion} from the residual update cancels $\hat{\mathbf{Q}}^l$ and yields the claim. Hence, the compression error introduced at layer $l^*$ passes through all subsequent layers unchanged: it cannot be reduced by any quantization that reuses $\mathbf{A}^\star$. \hfill$\blacksquare$

\noindent\textbf{Theorem 1 (Hierarchy of error bounds).}
After all $C$ quantization layers, the expected residual error satisfies:
\begin{align}
    \mathbb{E}[\|\mathbf{R}^{C+1}\|_F^2]
    \leq \Big[ 1 - \alpha_{l^*} \big( 1 - \epsilon_{\max}^{\,C - l^* + 1} \big) \Big]
    \prod_{l=1}^{l^*-1} \lambda_l\,
    \mathbb{E}[\|\mathbf{H}\|_F^2].
    \label{eq:main_bound_full}
\end{align}
\noindent\textit{Proof.} Decompose the final residual as $\mathbf{R}^{C+1} = (\mathbf{R}^{C+1} - \mathbf{A}^\star \mathbf{R}^{C+1}) + \mathbf{A}^\star \mathbf{R}^{C+1}$. The two parts are orthogonal by Lemma~3, and applying Lemma~5 repeatedly over layers $l^*, \dots, C$ replaces the first part with the compression error at layer $l^*$:
\begin{align}
    \mathbb{E}[\|\mathbf{R}^{C+1}\|_F^2]
    &= \mathbb{E}\big[\|\mathbf{R}^{l^*} - \mathbf{A}^\star \mathbf{R}^{l^*}\|_F^2\big] \nonumber \\
    &\quad + \mathbb{E}\big[\|\mathbf{A}^\star \mathbf{R}^{C+1}\|_F^2\big].
    \label{eq:final_decomposition}
\end{align}
The first term equals $(1-\alpha_{l^*})\,\mathbb{E}[\|\mathbf{R}^{l^*}\|_F^2]$ by Equation~\eqref{eq:pythagoras} and the definition of $\alpha_{l^*}$ in Equation~\eqref{eq:energy_ratio}.

For the second term, consider any layer $l \geq l^*$. These layers
share the same compression matrix ($\mathbf{A}^l = \mathbf{A}^\star$),
so Equation~\eqref{eq:insubspace_recursion} shows that
$\mathbf{A}^\star \mathbf{R}^{l+1}$ is exactly the quantization error of layer $l$. Since $\mathbf{A}^l = \mathbf{A}^\star$, the quantizer input $\mathbf{A}^l \mathbf{R}^l$ equals $\mathbf{A}^\star \mathbf{R}^l$, so Lemma~2 (Equation~\eqref{eq:rd_bound}) applies with input $\mathbf{A}^\star \mathbf{R}^l$ and bounds it:
\begin{align}
    \mathbb{E}[\|\mathbf{A}^\star \mathbf{R}^{l+1}\|_F^2]
    = \mathbb{E}[\|\mathbf{A}^\star \mathbf{R}^{l} - \hat{\mathbf{Q}}^l\|_F^2]
    \leq \epsilon_{\max}\, \mathbb{E}[\|\mathbf{A}^\star \mathbf{R}^{l}\|_F^2].
\end{align}
Applying this bound recursively over the $C - l^* + 1$ layers from $l^*$ to $C$, and then converting the compressed energy at layer $l^*$ to residual energy via Equation~\eqref{eq:energy_ratio}:
\begin{align}
    \mathbb{E}[\|\mathbf{A}^\star \mathbf{R}^{C+1}\|_F^2]
    &\leq \epsilon_{\max}^{\,C-l^*+1}\, \mathbb{E}[\|\mathbf{A}^\star \mathbf{R}^{l^*}\|_F^2] \nonumber \\
    &= \epsilon_{\max}^{\,C-l^*+1}\, \alpha_{l^*}\, \mathbb{E}[\|\mathbf{R}^{l^*}\|_F^2].
\end{align}
Summing the two terms gives $\big[1 - \alpha_{l^*}\big(1 - \epsilon_{\max}^{\,C-l^*+1}\big)\big]\, \mathbb{E}[\|\mathbf{R}^{l^*}\|_F^2]$. Bounding $\mathbb{E}[\|\mathbf{R}^{l^*}\|_F^2]$ by unrolling Lemma~4 over layers $1, \dots, l^*-1$ yields Equation~\eqref{eq:main_bound_full}. \hfill$\blacksquare$

Equation~\eqref{eq:main_bound_full} reveals a hierarchy of bounds as $C \to \infty$:
\begin{itemize}
    \item \emph{Shared segmentation boundary ($l^* = 1$).} All layers reuse the boundaries of the first layer, corresponding to single compression. The second term of Equation~\eqref{eq:final_decomposition} vanishes and the error converges to a constant floor:
    \begin{align}
        \lim_{C \to \infty} \mathbb{E}[\|\mathbf{R}^{C+1}\|_F^2]
        = (1 - \alpha_1)\, \mathbb{E}[\|\mathbf{H}\|_F^2],
    \end{align}
    which is exactly the compression error of the first layer and is strictly positive by Assumption~1.
    \item \emph{Boundary anchor ($1 < l^* < C$).} The error floor is reduced by the contraction of the layer-wise phase:
    \begin{align}
        \lim_{C \to \infty} \mathbb{E}[\|\mathbf{R}^{C+1}\|_F^2]
        \leq (1 - \alpha_{l^*}) \prod_{l=1}^{l^*-1} \lambda_l\,
        \mathbb{E}[\|\mathbf{H}\|_F^2].
    \end{align}
    \item \emph{Layer-wise compression without anchor (LACE).} Every layer computes new boundaries, so Lemma~4 applies at every layer:
    \begin{align}
        \mathbb{E}[\|\mathbf{R}^{C+1}\|_F^2]
        \leq \prod_{l=1}^{C} \lambda_l\, \mathbb{E}[\|\mathbf{H}\|_F^2]
        \leq \lambda_{\max}^{\,C}\, \mathbb{E}[\|\mathbf{H}\|_F^2],
    \end{align}
    where $\lambda_{\max} = 1 - \alpha_{\min}(1 - \epsilon_{\max}) < 1$, so the bound converges to zero as $C \to \infty$.
\end{itemize}
Therefore, recomputing boundaries inside the residual loop removes the constant error floor inherent to a shared segmentation boundary, while the boundary anchor interpolates between the two regimes.

\noindent\textbf{Remark.} This analysis assumes that layers $l > l^*$ reuse the segmentation boundary of layer $l^*$ exactly. In our implementation (Section III-C), these layers may instead select any subset of $\mathcal{B}^{\text{anchor}}$. The theorem corresponds to the special case where the full segmentation boundary of layer $l^*$ is reused.

\bibliographystyle{IEEEtran}
\bibliography{mybib}